\documentclass[twocolumn,showpacs,preprintnumbers,amsmath,amssymb,superscriptaddress]{revtex4-2}
\usepackage{ulem}
\usepackage[dvipdfmx]{epsfig}
\usepackage{color}
\usepackage{bm}
\begin{document}
\preprint{NITEP 204}
\title{Isospin-forbidden electric dipole transition of the 9.64 MeV state of $^{12}$C}
\author{Y. Suzuki}
\email{y.suzuki@emeritus.niigata-u.ac.jp}
\affiliation{Department of Physics, Niigata University, Niigata 950-2181, Japan}
\affiliation{RIKEN Nishina Center, Wako 351-0198, Japan}
\author{W. Horiuchi}
\email{whoriuchi@omu.ac.jp}
\affiliation{Department of Physics, Osaka Metropolitan University, Osaka 558-8585, Japan}
\affiliation{Nambu Yoichiro Institute of Theoretical and Experimental Physics (NITEP), Osaka Metropolitan University, Osaka 558-8585, Japan}
\affiliation{RIKEN Nishina Center, Wako 351-0198, Japan}
\affiliation{Department of Physics,
  Hokkaido University, Sapporo 060-0810, Japan}

\author{M. Kimura}
\email{masaaki.kimura@ribf.riken.jp}
\affiliation{RIKEN Nishina Center, Wako 351-0198, Japan}
\affiliation{Department of Physics,
  Hokkaido University, Sapporo 060-0810, Japan}
%\date{\today}

\begin{abstract}
\noindent
The electric dipole transition of the 3$^-$ state at 9.64 MeV of 
$^{12}$C to the 2$^+$ state at 4.44 MeV is speculated to play a key role in the triple-$\alpha$ reaction at high temperatures. A theoretical prediction 
of its transition width is a challenge to nuclear theory  because it belongs 
to a class of isospin-forbidden transitions. We extend 
a microscopic 3$\alpha$ cluster-model to include isospin 1 impurity components, and take into account both 
isovector and isoscalar electirc dipole operators. 
Several sets of $2^+$ and $3^-$ wave functions are generated
by solving a radius-constrained equation of motion with
the stochastic variational method, 
resulting in reproducing very well the electric quadrupole and octupole 
transition probabilities to the ground state. The electric dipole transition 
width is found to be 7--31 meV, 16 meV on the average,  and more than  
half of the width is contributed by the isospin mixing of  $\alpha$ particles.
\end{abstract}
\maketitle

\section{Introduction}

It is well-known that the triple-$\alpha$ reaction is a key reaction to produce the elements heavier than $^{12}$C. At low temperatures, it occurs through the $0^+$ Hoyle state at 7.65 MeV~\cite{salpeter52,hoyle54}. Its importance is numerically confirmed~\cite{ishikawa13,suno16}, in reasonable agreement with the $R$-matrix prediction~\cite{nacre99} at $T < 0.1$ GK. 

At higher temperatures, $T > 2$ GK, relevant to supernovae and X-ray bursts, the triple-$\alpha$ reaction via the $3^-$ state at 9.64 MeV of $^{12}$C is presumed to play an important role. To estimate its impact, the radiative decay of the $3^-$ state relative to its total width has been measured~\cite{crannell67,chamberlin74}, and an upper limit of $\Gamma_{\rm rad}/\Gamma_{\rm total} < 4.1 \times 10^{-7}$ was deduced~\cite{ajzenberg-selove90,kelley17}. Recently, two independent experiments have attempted to update the ratio, indicating $\Gamma_{\rm rad}/\Gamma_{\rm total} = 1.3^{+1.2}_{-1.1} \times 10^{-6}$~\cite{tsumura21} and $6.4\pm 5.1 \times 10^{-5}$~\cite{Cardella}, respectively. Both of them appear 
to be much larger than the previous upper limit, although the substantial uncertainties make it difficult to determine whether or not the $3^-$ state really contributes to the synthesis of $^{12}$C at high temperatures. Hence, a theoretical evaluation is necessary and important.

It should be noted that the radiative decay width from the $3^-$ state to the first $2^+$ state is the primary source of the uncertainty, because the total width ($\Gamma_{\rm total}=46\pm 3$ keV~\cite{kelley17} or $34\pm 5$ keV~\cite{ajzenberg-selove90}) is well constrained~\cite{note24} and the electric octupole ($E3$) decay width to the ground state ($0.31\pm 0.04$ meV~\cite{Crannel}) is rather small. A width due to the magnetic quadrupole transition is also expected to be small. The radiative decay from the $3^-$ state to the $2^+$ state should
therefore be dominated by an electric dipole ($E1$) transition. 
The decay is, however, hindered in the 
long-wavelength approximation, because both states are 
considered to be  good isospin zero states. 
It is thus necessary to go beyond the long-wavelength approximation and 
furthermore to take into account the breaking of isospin symmetry, which is a 
challenging task to nuclear theory.

The purpose of the present study is to estimate the $E1$ decay rate by 
assuming that the relevant states are all described well by a microscopic 
3$\alpha$ cluster-model. Many calculations have been performed in the 
cluster model using effective two-nucleon central forces. See, e.g., Refs.~\cite{Uegaki, Kamimura, Descouvemont87}. The binding energy, the spectrum of $^{12}$C, and some other observables  are reasonably well reproduced. However, describing the $\alpha$ cluster with $\phi_{\alpha}^{(0)}=(0s)^4$ harmonic-oscillator configuration fails in calculating the $E1$ transition probability  
because all the 3$\alpha$ configurations are isospin zero states. Let us clarify the point of the present study.  Up to the leading-order term beyond the long-wavelength approximation, the $E1$ operator acting on an $A$-nucleon system reads as~\cite{baye12} 
\begin{align}
&E_{1\mu}=E_{1\mu}({\rm IV})+E_{1\mu}({\rm IS}),\notag \\
&E_{1\mu}({\rm IV})=e\sum_{i\in {\rm proton}} {\cal Y}_{1\mu}({\bm r}_i- {\bm R}),\notag \\
&E_{1\mu}({\rm IS})=-e\frac{k^2}{10}\sum_{i=1}^A ({\bm r}_i-{\bm R})^2{\cal Y}_{1\mu}({\bm r}_i-{\bm R})\notag\\
&+\frac{e\hbar k}{8m_pc}\frac{2i}{\hbar} \sum_{i=1}^A 
{\cal Y}_{1\mu}({\bm r}_i- {\bm R})\, ({\bm r}_i- {\bm R}) \cdot ({\bm p}_i-\textstyle{\frac{1}{A}}{\bm P}).
\label{E1op}
\end{align}
Here, $\bm{r}_i$ and $\bm {p_i}$ are respectively the position coordinate and the momentum of the $i$th nucleon, 
  $k$ the wave number of the $E1$ transition, $m_p$ the proton mass, $\bm R$ and $\bm P$ are respectively the center-of-mass coordinate and 
the total momentum, and ${\cal Y}_{lm}(\bm r)$ is the solid spherical harmonics 
\begin{align}
{\cal Y}_{lm}(\bm r)=r^lY_{lm}(\hat{\bm r}),
\end{align} 
where $\hat{\bm r}$ stands for the polar and azimuthal angles of $\bm r$. 
$E_{1\mu}({\rm IV})$ is isovector and the leading term of the long-wavelength approximation. It has no contribution to an isospin zero nucleus.   Therefore, we have to evaluate the contribution of the isoscalar term, $E_{1\mu}({\rm IS})$, to the $E1$ matrix element, provided that we use the $3\alpha$ cluster-model for $^{12}$C. Though $E_{1\mu}({\rm IS})$ further contains a spin-dependent term~\cite{baye12}, we ignore it 
because $^{12}$C is described by the $3\alpha$ cluster-model with zero 
total spin. Another variant of expression for the isoscalar $E1$ 
operator is also discussed in Ref.~\cite{baye12}. 

There is the possibility, however, that $E_{1\mu}({\rm IV})$ receives non-vanishing 
contribution in so far as the relevant states of $^{12}$C contain 
isospin impurity components. We take into account the isospin mixing 
assuming that the $\alpha$ particle contains a small component of 
isospin 1 impurity configuration, $\phi^{(1)}_{\alpha}$,
\begin{align}
\phi_{\alpha}=\sqrt{1-\epsilon^2}\phi_{\alpha}^{(0)}+\epsilon \phi^{(1)}_{\alpha},
\label{alpha.wf}
\end{align}
as described in Ref.~\cite{suzuki21}. Contrary to $E_{1\mu}({\rm IV})$, 
$E_{1\mu}({\rm IS})$ gives nonzero contributions between the main isospin zero components of $^{12}$C. Both contributions of the isovector and isoscalar terms may compete with each other. 

Three states of $^{12}$C play a main role in the present study: the 
ground state, the 2$^+$ state at $E_{\rm x}\,=$ 4.44 MeV, and the 3$^-$ state 
at $E_{\rm x}\,=$ 9.64 MeV. Since the $E1$ transition matrix element is sensitive to the sizes of the relevant states, we obtain the wave functions of those states by taking into account not only the energies but 
also other physical observables sensitive to the sizes, that is, the point proton radius of the ground state, 
the electric quadrupole ($E2$) transition probability, $B(E2)$,   
from the $2^+$ state to the ground state, and the $E3$ transition probability, $B(E3)$, of the $3^-$ state to the ground state. 

Section~\ref{formulation} describes a way of constructing the relevant 
states and then explains how to include the isospin impurity components 
in evaluating the $E1$ transition matrix element. Section~\ref{sec.result} presents results of calculation for the ground state, the $2^+$ state,
and the $3^-$ state and discusses the isospin-forbidden $E1$ transition 
probability. A brief summary is drawn in Sec.~\ref{summary}.

\section{Formulation}
\label{formulation}

As is well-known, it is very hard to 
reproduce the binding energy and the excitation energies of the three states 
of $^{12}$C with effective nuclear forces such as Volkov~\cite{volkov65} and Minnesota~\cite{minnesota} potentials. 
Instead of minimizing the Hamiltonian expectation value, we constrain the 
size or radius of the system and study the energy as a function of the radius. This is reasonable because the size of the system is expected to play a vital role in the present study. 
We introduce a combination of operators, the Hamiltonian, $H$,
and the mean square radius, $R^2$,  
\begin{align}
S(\lambda)=H+\lambda R^2.
\label{lagrangian}
\end{align}
Here, $\lambda$ is a parameter. Given $\lambda$, we search for such a solution that the expectation value of $S(\lambda)$ becomes a minimum, denoted by $\left<S\right>_{\lambda}$. 
Using the obtained wave function, we evaluate the expectation values, $\left<H\right>$ and $\left<R^2\right>$. In this way we can study both energy and size of the 
relevant state at the same time as a function of $\lambda$. 

Except for the point proton radius of the ground state, there is no direct 
information on the sizes of the 2$^+$ state and the 3$^-$ resonance state. 
As noted above, however, we can make use of the $B(E2)$ 
value of the 2$^+$ state and the $B(E3)$ value of the 3$^-$ state to determine $\lambda$. 
We take  non-negative $\lambda$ in the present study. 
A negative value of $\lambda$ may play a role when $^{12}$C extends to strongly deformed configurations or fragments into $\alpha+2\alpha$ 
or $\alpha+\alpha+\alpha$ system.  In what follows, $\lambda$ is denoted by 
 $\lambda_L$, where $L$ is the total angular momentum of $^{12}$C.

The minimization of $S(\lambda_L)$ is performed by taking a 
combination of correlated Gaussian (CG) bases $\Phi_{LM}$~\cite{varga95}: 
\begin{align}
&\Psi_{LM}=\sum_{k=1}^K C(k) \Phi_{LM}(k),\notag \\
&\Phi_{LM}(k)={\cal A}\Big\{{\cal Y}_{LM}(\widetilde{u_k}\bm x)
e^{-\frac{1}{2}\tilde{\bm x}A_k \bm x}\prod_{i=1}^3\phi^{(0)}_{\alpha}(i)\Big\},
\label{cgbasis}
\end{align}
where ${\cal A}$ is the antisymmetrizer of 12 nucleons. 
The coordinate $\bm x$ is a 2-dimensional column vector 
specifying the relative coordinates of 3 $\alpha$-particles: ${\bm x}_1=
{\bm R}_1-{\bm R}_2, \ {\bm x}_2=\frac{1}{2}({\bm R}_1+{\bm R}_2)-{\bm R}_3$, where ${\bm R}_i$ stands for the center-of-mass coordinate of the $i$th 
$\alpha$-particle. The CG basis $\Phi_{LM}(k)$ is characterized by variational parameters, $u_k$ and $A_k$: $u_k$ is a column vector of 2-dimension, and $A_k$ is a $2\times 2$ real, symmetric, 
positive-definite matrix. The tilde symbol $\tilde{\ }$ stands for the 
transpose of a column vector, that is, 
 $\widetilde{u_k}\bm x=u_k(1)\bm x_1+u_k(2)\bm x_2,\, 
\tilde{\bm x}A_k \bm x=A_k(1,1)\bm x_1^{\,2}+2A_k(1,2)\bm x_1\cdot \bm x_2+A_k(2,2)
\bm x_2^{\,2}$. No generality is lost by assuming $\widetilde{u_k}u_k=1$.  
Each CG basis thus contains 3 parameters for $L=0$ and 4 parameters for $L=2, 3$. 
The matrix $A_k$ is conveniently defined through three relative distance parameters, $(d_k(12), d_k(13), d_k(23))$, by~\cite{cpc97,book}
\begin{align}
\tilde{\bm x}A_k \bm x=\sum_{j>i=1}^3 \Big(\frac{{\bm R}_i-{\bm R}_j}{d_k(ij)}\Big)^{\,2}.
\end{align} 
We note the following in choosing the set $\{d_k(ij)\}$: The root-mean-square 
(rms) radius of the center-of-masses of 3 $\alpha$-particles 
is defined by $\sqrt{\frac{1}{3}\sum_{i=1}^3({\bm R}_i-{\bm R})^2}=\frac{1}{3}\sqrt{\sum_{j>i=1}^3 ({\bm R}_i-{\bm R}_j)^2}$, where ${\bm R}=\frac{1}{3}\sum_{i=1}^3{\bm R}_i$. Therefore,   
$\bar{D}_k=\frac{1}{3}\sqrt{\sum_{j>i=1}^3d_k(ij)^2}$ controls the global size of the 3$\alpha$ system. We choose $\{d_k(ij)\}$ to 
make $\bar{D}_k$  cover sufficiently large values. 

Calculation of all the needed matrix elements can be done as explained in Ref.~\cite{suzuki21}. 
It is worthwhile to note that the angular-momentum projection is carried out analytically, which guarantees an accurate evaluation of all the matrix elements. 
This accuracy is a vitally important ingredient to make a stochastic search
of the basis set possible and practical.
Both $u_k$ and $A_k$ serve to control the 
partial-wave contents among $\alpha$-particles.  
The parameters, $u_k$ and $A_k$, are determined by the 
stochastic variational method~\cite{varga95,book}. The 
previous calculation for the $0^+$ case suggests that the basis dimension 
$K$ could be a small value~\cite{matsumura04}: $K$ is 
set to 7 for the 0$^+$ state and to 20 for the $2^+$ and $3^-$ states. 
The basis determination consists of (i) 
a trial and error search of the basis set 
up to $K$ dimension, followed by (ii) a refining search that 
replaces the already selected base with a new candidate base if the latter decreases 
the expectation value of $S(\lambda_L)$. Random bases tested in each step of 
 (i) and (ii) are typically 15--20. A refinement cycle is repeated more than ten times. It should be stressed that we have to obtain a well-converged 
solution to draw a reliable $B(E1)$ value because it could be 
very sensitive to the details 
of the relevant wave functions.

The wave function of  $\alpha$-particle $\phi^{(0)}_{\alpha}$ 
is constructed from the $(0s)^4$ harmonic-oscillator configuration 
with its center-of-mass motion excluded. Its single-particle orbit is 
a Gaussian function, ${\rm exp}(-\frac{\beta}{2}r^2)$, 
with $\beta=0.52$ fm$^{-2}$. Since $\epsilon$ is on the order of 
$10^{-3}$~\cite{suzuki21}, evaluating the matrix elements of 
$S(\lambda_L)$ is carried out with $\phi^{(0)}_{\alpha}$ as defined in 
Eq.~(\ref{cgbasis}) but not with $\phi_{\alpha}$ of 
Eq.~(\ref{alpha.wf}).  
The impurity component $\phi^{(1)}_{\alpha}$ is normalized and has the same 
quantum numbers as $\phi^{(0)}_{\alpha}$ except for the isospin. The 
spatial part of $\phi^{(1)}_{\alpha}$  is constructed from a 2$\hbar \omega$ excited shell-model configuration with its spurious center-of-mass motion being excluded~\cite{suzuki21}.  Once $\Psi_{LM}$ is 
obtained, it is reasonable to define a 3$\alpha$ wave function with isospin mixing, $\Psi^\prime_{LM}$, by replacing $\phi^{(0)}_{\alpha}(i)$ with $\phi_{\alpha}(i)$ in Eq.~(\ref{cgbasis}). 
Since $\epsilon$ is sufficiently small, $\Psi^\prime_{LM}$ can be very well approximated up to the first order in $\epsilon$ as follows: 
\begin{align}
\Psi^\prime_{LM} &= (1-\epsilon^2)^{\frac{3}{2}} \Psi_{LM} +\epsilon (1-\epsilon^2)\sum_{i=1}^3 \Psi_{LM}(i) +\cdots \notag \\
               & \approx \Psi_{LM} +\epsilon \sum_{i=1}^3 \Psi_{LM}(i).
\end{align}
Here, $\Psi_{LM}(i)$ is defined by replacing $\phi^{(0)}_{\alpha}(i)$ by $\phi^{(1)}_{\alpha}(i)$ in Eq.~(\ref{cgbasis}), whereas  
the rest of the $\alpha$-particle wave functions is unchanged, thus $\Psi_{LM}(i)$ has isospin 1. 
Since the $E1$ operator consists of the isovector  and isoscalar terms, Eq.~(\ref{E1op}),  the $E1$ matrix element read as  
\begin{align}
&\langle \Psi^\prime_{L^\prime M^\prime} |E_{1\mu}({\rm IV})+E_{1\mu}({\rm IS})|\Psi^\prime_{LM} \rangle\notag \\
  &\ \ \ \approx \langle \Psi_{L^\prime M^\prime}|E_{1\mu}({\rm IS})|\Psi_{LM}\rangle\notag\\
  &\ \ \ + \epsilon 
  \sum_{i=1}^3\Big\{  \langle \Psi_{L^\prime M^\prime}|E_{1\mu}({\rm IV})|\Psi_{LM}(i)\rangle\Big.\notag\\
  \Big.
&\ \ \ \ \ \ \ \ \ \ \ \ 
+\langle \Psi_{L^\prime M^\prime}(i)|E_{1\mu}({\rm IV})|\Psi_{LM} \rangle \Big\}.
\label{E1formula}
\end{align}
Here, the first term is the contribution of the isoscalar $E1$ operator between the main components 
of the wave functions with isospin 0, while the second terms are the contributions 
of the isovector $E1$ operator including the small components of the 3$\alpha$ wave function with isospin 1 either in the ket or in the bra.  

It should be noted that the transition matrix element $\langle \Psi_{L^\prime M^\prime}|E_{1\mu}({\rm IV})|\Psi_{LM}(i)\rangle$
can be nonzero 
only when $E_{1\mu}({\rm IV})|\Psi_{LM}(i)\rangle$  
contains the same spin-isospin functions as that of $\Psi_{L^\prime M^\prime}$, that is, 
a product of three totally antisymmetric spin-isospin functions. 
As was done in Ref.~\cite{suzuki21}, it is convenient to decompose 
$E_{1\mu}({\rm IV})$ into $E_{1\mu}({\rm IV})=\sum_{p=1}^3 E_{1\mu}({\rm IV},p)$ with  
\begin{align}
  E_{1\mu}({\rm IV},p)=e \sum_{q=1}^4 \Big(\frac{1}{2}-t_3(p_q)\Big)
  {\cal Y}_{1\mu}({\bm r}_{p_q} -{\bm R}_{p}),
\end{align}
where  $t_3$ is the $z$ component of the nucleon isospin, 
$p_q$ is introduced to 
denote the nucleon label of the $p$th $\alpha$-particle, and its 
center-of-mass coordinate is given by 
${\bm R}_p=\frac{1}{4}\sum_{q=1}^4 {\bm r}_{p_q}$.  Only the $p=i$ term among 
three  sums over $p$   
satisfies the condition, leading to  
\begin{align}
\sum_{i=1}^3 E_{1\mu}({\rm IV})|\Psi_{LM}(i)\rangle \rightarrow E_{1\mu}^{\rm eff}({\rm IS})|\Psi_{LM}\rangle,
\label{e1.eff}
\end{align}
where  $E_{1\mu}^{\rm eff}({\rm IS})$ is an effective isoscalar $E1$ operator 
given by 
\begin{align}
E_{1\mu}^{\rm eff}({\rm IS})=-e\frac{2\beta}{3\sqrt{3}} \sum_{i=1}^3 \sum_{j=1}^4 ({\bm r}_{i_j}-{\bm R}_i)^2 {\cal Y}_{1\mu}({\bm r}_{i_j}-{\bm R}_i).
\label{E1.eff.op}
\end{align}
Substituting Eq.~(\ref{e1.eff}) into Eq.~(\ref{E1formula}) enables us to evaluate the $E1$ matrix element including the effect of the isospin mixing as follows:
\begin{align}
  &\langle \Psi'_{L^\prime M^\prime} |E_{1\mu}({\rm IV})+E_{1\mu}({\rm IS})|\Psi'_{LM} \rangle\notag\\
  &\approx \langle \Psi_{L^\prime M^\prime}|E_{1\mu}({\rm IS})|\Psi_{LM}\rangle+ 2 \epsilon 
\langle \Psi_{L^\prime M^\prime}|E_{1\mu}^{\rm eff}({\rm IS})|\Psi_{LM}\rangle.
\label{e1final}
\end{align}  
The effect of the isospin mixing is thus taken care of in the conventional 
$\alpha$ cluster-model. What is needed is to calculate the matrix 
elements of $E_{1\mu}^{\rm eff}({\rm IS})$. It is interesting to compare the matrix elements of 
different types of isoscalar operators for the $E1$ transition from the $3^-$ state 
to the $2^+$ state.

\section{Results of calculation}
\label{sec.result}

\begin{table}[t]
\caption{Results for the $0^+$ ground state of $^{12}$C.   $\left<S\right>_{\lambda_0}$ and $\left<H\right>$ are values from 
3$\alpha$ threshold. The observed $0^+$ ground state is located at $-7.27$ MeV below $3\alpha$ threshold~\cite{ajzenberg-selove90,kelley17}, and the point rms radius is 2.33\,fm~\cite{angeli13}.}
\begin{tabular}{ccccccc}
\hline\hline
$\lambda_0$ && $\left<S\right>_{\lambda_0}$ && $\left<H\right>$ && $\sqrt{\left<R^2\right>}$  \\
\hline
MeV $\cdot$ fm$^{-2}$&& MeV   && MeV && fm \\
\hline
0.0   && $-$6.008 && $-$6.008 && 2.456 \\
0.4   && $-$3.642 && $-$5.959 && 2.406 \\
0.8   && $-$1.318 && $-$5.835 && 2.376 \\
1.2   && 0.903 && $-$5.717 && 2.349 \\
1.6  && 3.148  && $-$5.570 && 2.334 \\
2.0  && 5.354 && $-$5.356 && 2.314  \\
\hline\hline
\end{tabular}
\label{0+}
\end{table}
 
\begin{figure}[ht]
\begin{center}
  \epsfig{file=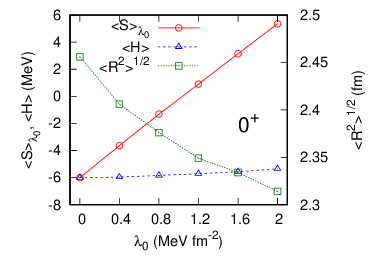, width=\columnwidth}
  \caption{$\langle S \rangle_{\lambda_0}$, $\langle H \rangle$, and 
    $ \langle R^2 \rangle$ as a function of $\lambda_0$.
 Thin lines are drawn for a guide to the eye.}
    \label{0+fig}
  \end{center}
\end{figure}

We use Volkov No.1 two-nucleon central potential~\cite{volkov65} with $m=0.59$, where $m$ is the parameter responsible 
for the Majorana exchange component.  The value around $m=0.6$ is consistent with 
$\alpha \alpha$ scattering data~\cite{alpha-alpha}. The two-nucleon Coulomb potential is 
included. The energy of $\alpha$-particle turns out to be $-27.076$\,MeV.  
Table~\ref{0+} lists $L=0$ results of converged solutions 
as a function of $\lambda_0$.    
The case of $\lambda_0=0$ is the usual energy minimization. Interestingly, the variation 
of $\left<S\right>_{\lambda_0}$ as a function of $\lambda_0$ is quite large compared to those of 
$\left<H\right>$ and $\left<R^2\right>$. See also Fig. ~\ref{0+fig}. As seen from the table, the $\lambda_0=0$ case predicts 
too large point rms radius 
for the ground state, which is about 2.33 fm~\cite{angeli13}.  
Instead of 
this usual approach, we determine the $0^+$ ground state 
to be such a solution that reproduces the rms radius. The appropriate 
value of $\lambda_0$ is found to be about 1.6 MeV$\cdot$fm$^{-2}$. In what 
follows, we set up the ground state to be the solution obtained with 
$\lambda_0$= 1.6 MeV$\cdot$fm$^{-2}$. Note that the ground state energy is then 
$-5.57$ MeV from the 3$\alpha$ threshold, which is about 1.7 MeV too high 
compared to experiment.

\begin{figure}[ht]
\begin{center}
  \epsfig{file=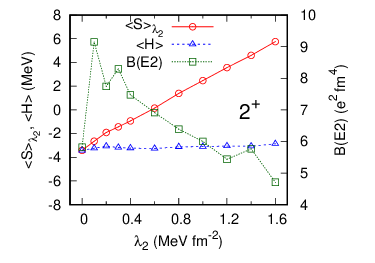, width=\columnwidth}
  \caption{$\langle S \rangle_{\lambda_2}$, $\langle H \rangle$, and 
    $B(E2)$ value as a function of $\lambda_2$.
 Thin lines are drawn for a guide to the eye.}
    \label{2+fig}
  \end{center}
\end{figure}

Table~\ref{2+} lists results of calculation for $L=2$. 
An appropriate value of $\lambda_2$ is determined by examining both the energy and the $B(E2)$ value to the ground state. The experimental values 
are respectively $-$2.84 MeV 
from the 3$\alpha$ threshold and $7.77\pm 0.43$ $e^2\,$fm$^4$~\cite{ajzenberg-selove90,kelley17}. Figure~\ref{2+fig} shows $\left<S\right>_{\lambda_2}$, $\left<H\right>$ and $B(E2)$ value as a function of $\lambda_2$.
The case with $\lambda_2=0$ 
predicts that the $2^+$ energy is lower than experiment by about 0.6 MeV and 
the $B(E2)$ value is smaller than experiment. With  $\lambda_2=0.2$--0.4 MeV$\cdot$fm$^{-2}$ both $\left<H\right>$ and $B(E2)$ become closer to the experimental values. 
The electric quadrupole moment $Q$  is calculated from $\sqrt{\frac{2}{35}}\langle \Psi_2\|Q_{\rm op}\|\Psi_2 \rangle$ without assuming an intrinsic 
shape, where the double barred matrix element stands for a reduced matrix element. 
Theory appears to give slightly smaller value than  $Q=6\pm 3$ $e\,$fm$^2$~\cite{vermeer83}.

\begin{table}[t]
\caption{Results for the first excited $2^+$ state of $^{12}$C located at $-2.84$ MeV below 3$\alpha$ threshold~\cite{ajzenberg-selove90,kelley17}.  $Q$ is the electric quadrupole moment. The 
$B(E2)$ value extracted from the radiative width $\Gamma_{\gamma}$~\cite{ajzenberg-selove90,kelley17} is $7.77\pm 0.43$ $e^2\,$fm$^4$.} 
\begin{tabular}{ccccccccccc}
\hline\hline
$\lambda_2$  && $\left<S\right>_{\lambda_2}$ && $\left<H\right>$ && $\sqrt{\left<R^2\right>}$ && $B(E2)$ && $Q$ \\
\hline
MeV$\cdot$fm$^{-2}$ && MeV && MeV && fm  && $e^2\,$fm$^4$ && $e\,$fm$^2$ \\
\hline
0.0  && $-$3.418 && $-$3.418 && 2.415  && 5.824 && $-$1.856 \\
0.1 && $-$2.643  && $-$3.238 && 2.438  && 9.150 && 1.441 \\
0.2   && $-$1.905 && $-$3.068 && 2.411 && 7.744 && $-$0.255 \\
0.3  && $-$1.431 && $-$3.173 && 2.410  && 8.299 && 1.021 \\
0.4   && $-$0.929 && $-$3.228 && 2.397 && 7.475 && 0.569 \\
0.6  && 0.147   && $-$3.261 && 2.383  && 6.914 && 1.410 \\
0.8   && 1.390 && $-$3.118 && 2.374 && 6.389 && 1.406  \\
1.0   && 2.470 && $-$3.101 && 2.360 && 6.000 && 1.251 \\
1.2   && 3.570 && $-$3.048 && 2.348 && 5.438 && $-$0.032 \\
1.4  && 4.596 && $-$3.072 && 2.340  && 5.765 && 1.906 \\
1.6  && 5.759 && $-$2.866 && 2.322  && 4.710 && $-$0.309 \\
\hline\hline
\end{tabular}
\label{2+}
\end{table}

\begin{table}[h]
\caption{Results for the $3^-$ resonance state of $^{12}$C located 2.37 MeV above $3\alpha$ threshold~\cite{ajzenberg-selove90,kelley17}. The $B(E3)$ value extracted from the radiative width $\Gamma_{\gamma_0}$~\cite{kelley17} is $107\pm 14$ $e^2\,$fm$^6$. }
\begin{tabular}{ccccccccc}
\hline\hline
$\lambda_3$ && $\left<S\right>_{\lambda_3}$ && $\left<H\right>$ && $\sqrt{\left<R^2\right>}$  && $B(E3)$  \\
\hline
MeV$\cdot$fm$^{-2}$&& MeV  && MeV && fm  && $e^2\,$fm$^6$  \\
\hline
0.4   && 4.703 && 1.626 && 2.773 && 378.7  \\
0.6  && 6.186 && 1.775 && 2.711  && 636.9  \\
0.7   && 6.900 && 1.840 && 2.689 && 6.331  \\
0.8  && 7.621 && 1.928 && 2.668  && 22.14  \\
0.9  && 8.333 && 2.004 && 2.652  && 78.16  \\
1.0   && 9.034 && 2.093 && 2.635 && 126.9  \\
1.1   && 9.722 && 2.169 && 2.620 && 104.4  \\
1.2  && 10.403 && 2.248 && 2.607  && 92.99  \\
1.3  && 11.078 && 2.327 && 2.595  && 95.50  \\
1.4   && 11.754 && 2.405 && 2.584 && 85.51  \\
1.5   && 12.416  && 2.474 && 2.575&& 44.61  \\
1.6 && 13.067  && 2.544 && 2.565  && 44.18  \\
\hline\hline
\end{tabular}
\label{3-}
\end{table}

\begin{figure}[ht]
\begin{center}
  \epsfig{file=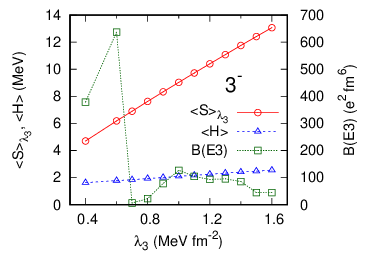, width=\columnwidth}
  \caption{$\langle S \rangle_{\lambda_3}$, $\langle H \rangle$, and 
    $B(E3)$ value as a function of $\lambda_3$.
 Thin lines are drawn for a guide to the eye.}
    \label{3-fig}
  \end{center}
\end{figure}

Table~\ref{3-} presents results of calculation for $L=3$. 
The $3^-$ state of $^{12}$C is by 2.37 MeV above $3\alpha$ threshold with the  
total width of $46\pm 3$ keV~\cite{kelley17}. Its radiative decay width 
is less than 19 meV, and its partial width to the ground state decay 
is $(3.1\pm0.4)\times 10^{-4}\,$eV~\cite{ajzenberg-selove90,kelley17}, indicating $B(E3: 3^- \to 0^+)=107\pm14\ e^2\,$fm$^{6}$. As the total width of the 3$^-$ state  is considerably small, 
it appears reasonable to treat the state as a bound state.    
The case of $\lambda_3=0$ cannot lead to a bound-state solution as expected:  $\left<H\right>$ 
tends to be zero and $\sqrt{\left<R^2\right>}$ becomes larger and larger as the basis set 
reaches large distances. With positive $\lambda_3$, however, we obtain a positive-energy bound-solution as listed in Table~\ref{3-}. Figure~\ref{3-fig} displays $\left<S\right>_{\lambda_3}$, $\left<H\right>$, and $B(E3)$ value as a function of $\lambda_3$.
The observed excitation energy and the $B(E3)$ value 
are fairly well reproduced by taking $\lambda_3$ in the range of 1.0--1.4 MeV$\cdot$fm$^{-2}$. 

\begin{table}[ht]
\caption{Radiation width of the $3^-$ resonance state of $^{12}$C to 
the 2$^+$ state due to the electric dipole transition. The width, in units of 
meV, is calculated 
without the isospin mixing ($\epsilon=0$) or with the isospin mixing. 
The upper limit of the total radiation width of the $3^-$ resonance state is 19 meV~\cite{kelley17}. Values in parentheses, in units of $10^{-2}e\,$fm, denote the $E1$ reduced matrix 
elements contributed by three kinds of isoscalar $E1$ operators. See Eq.~(\ref{E1content}). 
}
\begin{tabular}{cccccc}
\hline\hline
$\lambda_3$  && $\lambda_2$ &&\multicolumn{2}{l}{$\quad \Gamma_{\rm rad}(3^- \to 2^+)$}\\
\cline{4-6}
     &&     && $\epsilon=0$ & $\epsilon=-0.0042$     \\
\hline
1.0  && 0.2 && 8.58         & 26.0   \\
     &&     &&              &(7.41, $-$5.39, 1.50) \\
1.0  && 0.4 && 6.99         & 11.7   \\
     &&     &&              &(7.21, $-$5.38, $-$4.18) \\
1.1  && 0.2 && 3.34         & 9.49  \\
     &&     &&              & (5.01,$-$3.75, 0.86) \\
1.2  && 0.2 && 3.15         & 7.51  \\
     &&     &&              &(4.83, $-$3.61, 0.67) \\
1.2  && 0.4 && 2.44         & 19.2   \\
     &&     &&              &(4.70, $-$3.62, $-$4.10) \\
1.4  && 0.2 && 2.34         & 6.52   \\
     &&     &&              &(4.86, $-$3.81, 0.71) \\
1.4  && 0.4 && 2.21         & 30.7   \\
     &&     &&              &(4.85, $-$3.83, $-$4.84) \\
\hline\hline
\end{tabular}
\label{gamrad}
\end{table}

Using the $2^+$ and $3^-$ wave functions obtained above, we evaluate the 
$B(E1)$ value. The isospin mixing parameter $\epsilon$ 
is set to $-4.2\times 10^{-3}$~\cite{suzuki21}. The radiation 
width due to the $E1$ transition is calculated from 
\begin{align}
\Gamma_{\rm rad}(3^- \to 2^+)= 1.473 \times 10^5 B(E1: 3^- \to 2^+),
\end{align}
where $B(E1: 3^- \to 2^+)$ is given in units of $e^2{\, \rm fm}^2$ 
and $\Gamma_{\rm rad}$ is in units of meV.
The `best' wave functions obtained with $\lambda_3=1.1$ and $\lambda_2=0.2$ MeV$\cdot$fm$^{-2}$
predict the $B(E3)$ and $B(E2)$ values closest to the respective medians
of the experimental values. In addition to this case,
we test three $3^-$ states obtained 
with $\lambda_3=1.0, 1.2$, and 1.4\,MeV$\cdot$fm$^{-2}$
together with two $2^+$ states calculated from 
$\lambda_2=0.2$ and 0.4\,MeV$\cdot$fm$^{-2}$. The $E1$ radiation width calculated from a combination of these wave functions is listed in 
Table~\ref{gamrad}.  The largest width among 7 cases 
is 31 meV, the width 
from the `best' combination is 9.5 meV, and the average of the widths 
is 16 meV consistently with the upper limit of $\Gamma_{\rm rad} < 19$ meV~\cite{kelley17}. The ratio of $\Gamma_{\rm rad}(3^- \to 2^+)/\Gamma_{\rm total}(3^-)$ with $\Gamma_{\rm total}(3^-)=46\pm 3$ keV turns out to be 0.35$\times 10^{-6}$ on the average.  The corresponding ratio of Ref.~\cite{tsumura21} is $1.3^{+1.2}_{-1.1} \times 10^{-6}$. The calculated theoretical ratio is within the error bars of 
that experiment, but it is much smaller than that quoted in Ref.~\cite{Cardella}. If the isospin mixing of $\alpha$ particles is not taken into account, the 
$\Gamma_{\rm rad}(3^- \to 2^+)$ value decreases to 4 meV on the average. 

The average value of $\Gamma_{\rm rad}=16$ meV corresponds to $3.2\times 10^{-4}$ W.u. It is interesting to compare this value to the case of $^{16}$O 
where the $E1$ transition  is isospin-forbidden and $\Gamma_{\rm rad}(E1)$ values are known. The $\Gamma_{\rm rad}(E1)$ values 
in Weisskopf units are $(3.5\pm 0.4)\times 10^{-4}$ for the $1^{-}$ state at 
7.12 MeV and $(6.0\pm0.9)\times 10^{-5}$ for the $1^-$ state at 9.63 MeV, 
respectively~\cite{ajzenberg-selove86}. The value we obtain for $^{12}$C
is in good correspondence with the $^{16}$O case.
This indicates that the approach developed in this paper
is sound and useful for evaluating
the isospin-forbidden $E1$ transition strength.

Also shown in the table is the contribution of each of three different 
isoscalar operators to the $B(E1)$ value, 
\begin{align}
B(E1: 3^- \to 2^+)=\frac{1}{7}\Big|\sum_{i=1}^3 \langle \Psi_{2}\| E1({\rm IS};i)\|\Psi_{3} \rangle\Big|^2.
\label{E1content}
\end{align}
Here, three kinds of isoscalar $E1$ operators are 
 (i) the first term of $E_{1\mu}({\rm IS})$ in Eq.~(\ref{E1op}), 
 (ii) the second term of $E_{1\mu}({\rm IS})$ in Eq.~(\ref{E1op}), and 
 (iii) the effective isoscalar $E1$ operator, Eq.~(\ref{E1.eff.op}).
The reduced matrix element of the first type is larger in its magnitude than that of the 
second type and they cancell each other. The isospin 
impurity term appears to increase the $B(E1)$ value in most cases, but its 
magnitude fluctuates depending on the choice of $\lambda_2$ and $\lambda_3$. 

It is convenient to express the reduced matrix 
element as a product of a numerical constant, $CC(i)$, and an integral,
${\rm RME}(i)$, as follows:
\begin{align}
\langle \Psi_{2}\|E1({\rm IS};i)\|\Psi_{3} \rangle= CC(i)\, {\rm RME}(i),
\end{align}
where
\begin{align}
  &CC(1)=-\frac{k^2}{10},\quad CC(2)=\frac{\hbar k}{4m_pc R^2},
  \quad CC(3)=-\frac{4\beta \epsilon}{3\sqrt{3}},\notag \\
&{\rm RME}(1)=e\, \langle r^2{\cal Y}_1(\bm r) \rangle,
\quad {\rm RME}(2)=e\, R^2 \langle {\bm r}\cdot {\bm \nabla}{\cal Y}_1(\bm r) \rangle,\notag\\
        &{\rm RME}(3)=e\, \langle r^2{\cal Y}_1(\bm r) \rangle^\prime.
\end{align}
Here, $R$ is a radius introduced to make ${\rm RME}(2)$ have $e$ times length dimension, 
and the prime put for RME(3) is to stress that the operator 
involved there is not the same as that of RME(1). It is very likely that $|$RME(3)$|$ is smaller than $|$RME(1)$|$. The operators involved in RME(1) and 
RME(3) are both $r^2{\cal Y}_1(\bm r)$ type, but their ranges are different. 
In RME(1) $\bm r$ denotes the distance vector of each nucleon from the 
center-of-mass of $^{12}$C, whereas it is the distance vector between 
the nucleon and the $\alpha$-particle to which the nucleon belongs. The latter 
is short-ranged and it appears that its matrix element 
depends on more detailed properties of the wave function.

The importance of these 
three terms apparently depends on the wave number $k$ as well as 
the maginitude of RME($i$). Note that $CC(3)$ is a constant determined by the 
property of the $\alpha$ particle, whereas the other two depend on the 
$E1$ transition energy, that is, the wave number $k$. With the use of $k=0.0264$\, fm$^{-1}$, $\beta=0.52$\, fm$^{-2}$, $\epsilon=-0.0042$, and $R=2.52$\, fm, the relative ratio of $CC(i)$s' is 
\begin{align}
  CC(2)/CC(1)&=-\frac{0.525}{R^2 k}=-3.13,\notag \\
  CC(3)/CC(1)&=-\frac{0.0168}{k^2}=-24.1.
\end{align}
These ratios qualitatively explain the results of 
Table~\ref{gamrad}.  With decreasing $k$, the terms with $i=2, 3$ become more 
important.

\section{Summary}
\label{summary}

We have studied the electric dipole transition of the 9.64 MeV 3$^-$ state 
of $^{12}$C to the 4.44 MeV 2$^+$ state. The transition belongs to a class 
of isospin-forbidden transitions, demanding a study 
beyond the usual long-wavelength approximation of the electric transition 
operators. We have employed 
a microscopic $3\alpha$ cluster-model to generate the ground state, 
the $2^+$ state, and the $3^-$ state. In determining the 
wave functions of those states, however, we have attempted to reproduce 
experimental observables sensitive to their sizes in addition to 
their energies. 

We have used the stochastic variational method to determine the wave functions. 
Among several combinations of $2^+$ and $3^-$ wave functions 
obtained within the 
accuracy of the experimental observables, we have selected 
several candidates to estimate the electric dipole transition probability. 
We have taken into account 
not only the next-order term beyond the long-wavelength approximation but  
also isospin mixings in both states of $^{12}$C.  The 
resulting  $\Gamma_{\rm rad}(3^- \to 2^+)$ value ranges 7 to 31 meV, 
the average of those values is  16 meV, and  
more than half of the width is contributed by the isospin 
mixing of $\alpha$ particles. The $\Gamma_{\rm rad}$ value obtained here
is considerably larger than 2 meV that was assumed in Ref.~\cite{nacre99}.

This study has been motivated by a question of whether or not the 9.64 MeV 
state plays an important role to triple-$\alpha$ reactions at high 
temperatures. There is no experimental information at present to test 
the $\Gamma_{\rm rad}$ value reported here. However, it is well-known that 
the $E1$ transition of the 7.12 MeV $1^-$ state of $^{16}$O to its ground state 
plays a crucially important role in $^{12}$C$(\alpha, \gamma)^{16}$O 
radiative capture reactions near the Gamow window. The $E1$ transition in that case  is 
again isospin-forbidden. A study similar to the present 
one will be interesting and useful.  Furthermore, such calculation 
can directly be compared to the observed radiation width.

\acknowledgments
We would like to thank P. Descouvemont, T. Kawabata, and M. Matsuo
for useful communications. 
This work is in part supported by JSPS KAKENHI Grants Nos. 18K03635 and 22H01214.

\end{document}